\documentclass[aps,prb,twocolumn,superscriptaddress,showpacs]{revtex4}

\usepackage{graphicx}
\usepackage{dcolumn}
\usepackage{amsmath}
\usepackage{amssymb}

\begin{document}

\title{Interplay between superconductivity and pseudogap state in bilayer cuprate superconductors}

\author{Yu Lan}
\affiliation{Department of Physics and Siyuan Laboratory, Jinan University, Guangzhou 510632, China}

\author{Jihong Qin}
\affiliation{Department of Physics, University of Science and Technology Beijing, Beijing 100083, China}

\author{Shiping Feng}
\affiliation{Department of Physics, Beijing Normal University, Beijing 100875, China}

\begin{abstract}
The interplay between the superconducting gap and normal-state pseudogap in the bilayer cuprate superconductors is
studied based on the kinetic energy driven superconducting mechanism. It is shown that the charge carrier interaction
directly from the {\it interlayer} coherent hopping in the kinetic energy by exchanging spin excitations does not
provide the contribution to the normal-state pseudogap in the particle-hole channel and superconducting gap in the
particle-particle channel, while only the charge carrier interaction directly from the {\it intralayer} hopping in
the kinetic energy by exchanging spin excitations induces the normal-state pseudogap in the particle-hole channel and
superconducting gap in the particle-particle channel, and then the two-gap behavior is a universal feature for the
single layer and bilayer cuprate superconductors.
\end{abstract}

\pacs{74.20.Mn, 74.25.Dw, 74.72.Kf, 74.62.-c\\
Keywords: Superconducting gap; Pseudogap; Two-gap feature; Bilayer cuprate superconductor}

\maketitle

The conventional superconductors are characterized by the energy gap, which exists in the excitation spectrum below
the superconducting (SC) transition temperature $T_{\rm c}$, and therefore is corresponding to the energy for breaking
a Cooper pair of the charge carriers and creating two quasiparticles \cite{schrieffer83}. However, in the cuprate
superconductors, an energy gap called the normal-state pseudogap exists \cite{Hufner08,Timusk99} above $T_{\rm c}$ but
below the pseudogap crossover temperature $T^{*}$, which is associated with some anomalous properties.
Although the charge carrier pair gap in the cuprate superconductors has a domelike shape of the doping dependence
\cite{damascelli03,campuzano04}, the magnitude of the normal-state pseudogap is much larger than that of the charge
carrier pair gap in the underdoped regime \cite{Hufner08,Timusk99}, then it smoothly decreases upon increasing doping,
and seems to merge with the charge carrier pair gap in the overdoped regime, eventually disappearing together with
superconductivity at the end of the SC dome \cite{Hufner08}. In this case, the charge carrier pair gap and normal-state
pseudogap are thus two fundamental parameters of the cuprate superconductors whose variation as a function of doping
and temperature provides important information crucial to understanding the details of superconductivity
\cite{Hufner08,Timusk99}.

Experimentally, a large body of experimental data obtained by using different measurement techniques have provided
rather detailed information on the low-energy excitations of the single layer and bilayer cuprate superconductors
\cite{Hufner08,Timusk99,damascelli03,campuzano04}, where the Bogoliubov-quasiparticle nature of the low-energy
excitations is unambiguously established \cite{campuzano03}. However, there are numerous anomalies for the bilayer
cuprate superconductors \cite{damascelli03,campuzano04}, which complicate the physical properties of the low-energy
excitations in the bilayer cuprate superconductors. This follows a fact that the bilayer splitting (BS) has been
observed in the bilayer cuprate superconductors in a wide doping range \cite{dfeng01}, which derives the low-energy
excitation spectrum into the bonding and antibonding components due to the presence of the bilayer blocks in the unit
cell. In particular, it has been argued that this BS may play an important role in the form of the well pronounced
peak-dip-hump structure in the low-energy excitation spectrum of the bilayer cuprate superconductors
\cite{lan07,kordyuk02,borisenko03}. In this case, an important issue is whether the behavior of the normal-state
pseudogap observed in the low-energy excitation spectrum as a suppression of the spectral weight is universal or not.
Within the framework of the kinetic energy driven SC mechanism \cite{feng0306}, the interplay between the SC gap and
normal-state pseudogap in the single layer cuprate superconductors has been studied recently \cite{feng12}, where the
interaction between charge carriers and spins directly from the kinetic energy by exchanging spin excitations induces
the normal-state pseudogap state in the particle-hole channel and SC-state in the particle-particle channel, then there
is a coexistence of the SC gap and normal-state pseudogap in the whole SC dome. In particular, this normal-state
pseudogap is closely related to the quasiparticle coherent weight, and both the normal-state pseudogap and SC gap are
dominated by one energy scale. In this paper, we study the interplay between the SC gap and normal-state pseudogap in
the bilayer cuprate superconductors along with this line. We show explicitly that the weak charge carrier interaction
directly from the {\it interlayer} coherent hopping in the kinetic energy by exchanging spin excitations does not
provide the contribution to the normal-state pseudogap in the particle-hole channel and SC gap in the particle-particle
channel, while only the strong charge carrier interaction directly from the {\it intralayer} hopping in the kinetic
energy by exchanging spin excitations induces the normal-state pseudogap in the particle-hole channel and SC gap in the
particle-particle channel, and then the two-gap behavior is a universal feature for the single layer and bilayer
cuprate superconductors.

The single common feature in the layered crystal structure of the cuprate superconductors is the presence of the
two-dimensional CuO$_{2}$ plane \cite{damascelli03}, and then it is believed that the unconventional physics properties
of the cuprate superconductors is closely related to the doped CuO$_{2}$ planes \cite{anderson87}. In this case, it is
commonly accepted that the essential physics of the doped CuO$_{2}$ plane \cite{anderson87} is captured by the $t$-$J$
model on a square lattice. However, for discussions of the interplay between the SC gap and normal-state pseudogap in
the bilayer cuprate superconductors, the $t$-$J$ model can be extended by including the bilayer interaction as
\cite{lan07},
\begin{eqnarray}\label{tJmodel}
H&=&-t\sum_{i\hat{\eta}a\sigma}C^{\dagger}_{ia\sigma}C_{i+\hat{\eta}a\sigma}+t'\sum_{i\hat{\tau}a\sigma}
C^{\dagger}_{ia\sigma}C_{i+\hat{\tau}a\sigma}\nonumber\\
&&-\sum_{i\sigma}t_{\perp}(i)(C^{\dagger}_{i1\sigma}C_{i2\sigma}+H.c.)
+\mu\sum_{ia\sigma}C^{\dagger}_{ia\sigma}C_{ia\sigma}\nonumber\\
&&+J\sum_{i\hat{\eta}a}{\bf S}_{ia} \cdot {\bf S}_{i+\hat{\eta}a}
+J_{\perp}\sum_{i}{\bf S}_{i1} \cdot {\bf S}_{i2},
\end{eqnarray}
supplemented by the local constraint $\sum_{\sigma}C_{ia\sigma}^{\dagger}C_{ia\sigma}\leq 1$ to remove double
occupancy, where $a=1,2$ is plane index, the summation within the plane is over all sites $i$, and for each $i$, over
its nearest-neighbors $\hat{\eta}$ or the next nearest-neighbors $\hat{\tau}$, $C^{\dagger}_{ia\sigma}$ and
$C_{ia\sigma}$ are electron operators that respectively create and annihilate electrons with spin $\sigma$,
${\bf S}_{i}=(S^{x}_{i},S^{y}_{i},S^{z}_{i})$ are spin operators, $\mu$ is the chemical potential, while the interlayer
hopping has the form in the momentum space,
\begin{eqnarray}\label{interlayer}
t_{\perp}({\bf k})={t_{\perp}\over 4}(\cos k_{x} -\cos k_{y})^{2},
\end{eqnarray}
which describes coherent hopping between the CuO$_{2}$ planes. This functional form of the interlayer hopping
(\ref{interlayer}) is predicted on the basis of the local density approximation calculations \cite{chakarvarty95}, and
later the experimental observed BS agrees well with it \cite{dfeng01}. In this bilayer $t$-$J$ model (\ref{tJmodel}),
the crucial requirement is to impose the electron single occupancy local constraint, which can be treated properly in
analytical calculations within the charge-spin separation (CSS) fermion-spin theory \cite{feng04,feng08}, where the
constrained electron operators are decoupled as $C_{ia\uparrow}=h^{\dagger}_{ia\uparrow}S^{-}_{ia}$ and
$C_{ia\downarrow}=h^{\dagger}_{ia\downarrow}S^{+}_{ia}$, with the spinful fermion operator
$h_{ia\sigma}= e^{-i\Phi_{ia\sigma}}h_{ia}$ that represents the charge degree of freedom together with some effects of
the spin configuration rearrangements due to the presence of the doped hole itself (charge carrier), while the spin
operator $S_{ia}$ describes the spin degree of freedom, then the electron single occupancy local constraint is
satisfied in analytical calculations. In this CSS fermion-spin representation, the bilayer $t$-$J$ model
(\ref{tJmodel}) can be expressed as,
\begin{eqnarray}\label{csstJmodel}
H&=&t\sum_{i\hat{\eta}a}(h^{\dagger}_{i+\hat{\eta}a\uparrow}h_{ia\uparrow}S^{+}_{ia}S^{-}_{i+\hat{\eta}a}+
h^{\dagger}_{i+\hat{\eta}a\downarrow}h_{ia\downarrow}S^{-}_{ia}S^{+}_{i+\hat{\eta}a})\nonumber\\
&&-t'\sum_{i\hat{\tau}a}(h^{\dagger}_{i+\hat{\tau}a\uparrow}h_{ia\uparrow}S^{+}_{ia}S^{-}_{i+\hat{\tau}a}
+h^{\dagger}_{i+\hat{\tau}a\downarrow}h_{ia\downarrow}S^{-}_{ia}S^{+}_{i+\hat{\tau}a})\nonumber \\
&&+\sum_{i}t_{\perp}(i)(h^{\dagger}_{i2\uparrow}h_{i1\uparrow}S^{+}_{i1}S^{-}_{i2}+h^{\dagger}_{i1\uparrow}
h_{i2\uparrow}S^{+}_{i2}S^{-}_{i1}\nonumber\\
&&+h^{\dagger}_{i2\downarrow}h_{i1\downarrow}S^{-}_{i1}S^{+}_{i2}
+h^{\dagger}_{i1\downarrow}h_{i2\downarrow}S^{-}_{i2}S^{+}_{i1})-\mu\sum_{ia\sigma}h^{\dagger}_{ia\sigma}h_{ia\sigma}\nonumber \\
&&+{J_{\rm eff}}\sum_{i\hat{\eta}a}{\bf S}_{ia}\cdot
{\bf S}_{i+\hat{\eta}a}+{J_{\rm eff\perp}}\sum_{i}{\bf S}_{i1}\cdot {\bf S}_{i2},
\end{eqnarray}
where $J_{\rm eff}=J(1-\delta)^{2}$, $J_{\rm eff\perp}=J_{\perp}(1-\delta)^{2}$, and
$\delta=\langle h^{\dagger}_{ia\sigma}h_{ia\sigma}\rangle=\langle h^{\dagger}_{ia} h_{ia}\rangle$
is the doping concentration.

For the bilayer cuprate superconductors, there are two coupled CuO$_{2}$ planes in one unit cell. In this case, the SC
order parameter for the electron Cooper pair is a matrix \cite{lan07} $\Delta=\Delta_{\rm L}+\sigma_{x}\Delta_{\rm T}$,
with $\Delta_{\rm L}$ and $\Delta_{\rm T}$ are the corresponding longitudinal and transverse parts, respectively. In
the doped regime without an antiferromagnetic long-range order (AFLRO), the charge carriers move in the background of
the disordered spin liquid state, and then the longitudinal and transverse SC order parameters can be expressed in the
CSS fermion-spin representation as, $\Delta_{\rm L}=-\chi_{1}\Delta_{\rm hL}$ and
$\Delta_{\rm T}=-\chi_{\perp}\Delta_{\rm hT}$, with
\begin{subequations}\label{CSSpair}
\begin{eqnarray}
\Delta_{\rm hL}&=&\langle h_{i+\hat{\eta}a\downarrow}h_{ia\uparrow}-h_{i+\hat{\eta}a\uparrow}h_{ia\downarrow}\rangle,\\
\Delta_{\rm hT}&=&\langle h_{i2\downarrow}h_{i1\uparrow}-h_{i2\uparrow}h_{i1\downarrow}\rangle,
\end{eqnarray}
\end{subequations}
are the corresponding longitudinal and transverse parts of the charge carrier pair gap parameter, respectively, and
the spin correlation functions
$\langle S^{+}_{ia}S^{-}_{i+\hat{\eta}a}\rangle=\langle S^{-}_{ia}S^{+}_{i+\hat{\eta}a}\rangle=\chi_{1}$ and
$\langle S^{+}_{i1}S^{-}_{i2}\rangle=\langle S^{-}_{i1}S^{+}_{i2}\rangle=\chi_{\perp}$. The result in
Eq. (\ref{CSSpair}) shows that as in the single layer case \cite{feng0306}, the SC gap parameter in the bilayer cuprate
superconductors is also closely related to the corresponding charge carrier pair gap parameter, and therefore the
essential physics in the SC-state is dominated by the corresponding one in the charge carrier pairing state.

Within the framework of the kinetic energy driven SC mechanism \cite{feng0306}, the electronic structure of the bilayer
cuprate superconductors has been discussed \cite{lan07,feng08}, and the result shows that the low-energy excitation
spectrum is split into the bonding and antibonding components due to the presence of BS, then the observed
peak-dip-hump structure is mainly caused by BS, with the peak being related to the antibonding component, and the hump
being formed by the bonding component. Following our previous discussions \cite{lan07,feng08}, the self-consistent
equations that satisfied by the full charge carrier normal and anomalous Green's functions are obtained as,
\begin{subequations}\label{EliashbergEq}
\begin{eqnarray}
g({\bf k},\omega)&=&g^{(0)}({\bf k},\omega)+g^{(0)}({\bf k},\omega)[\Sigma^{(\rm h)}_{1}({\bf k},\omega)
g({\bf k},\omega)\nonumber\\
&&-\Sigma^{(\rm h)}_{2}(-{\bf k},-\omega)\Im^{\dagger}({\bf k},\omega)], \\
\Im^{\dagger}({\bf k},\omega)&=&g^{(0)}(-{\bf k},-\omega)[\Sigma^{(\rm h)}_{1}(-{\bf k},-\omega)
\Im^{\dagger}(-{\bf k},-\omega)\nonumber\\
&&+\Sigma^{(\rm h)}_{2}(-{\bf k},-\omega)g({\bf k},\omega)],
\end{eqnarray}
\end{subequations}
respectively, where the full charge carrier normal Green's function
$g({\bf k},\omega)=g_{\rm L}({\bf k},\omega)+\sigma_{x}g_{\rm T}({\bf k},\omega)$, the full charge carrier anomalous
Green's function
$\Im^{\dagger}({\bf k},\omega)=\Im^{\dagger}_{\rm L}({\bf k},\omega)+\sigma_{x}\Im^{\dagger}_{\rm T}({\bf k},\omega)$,
the charge carrier self-energies $\Sigma^{(\rm h)}_{1}({\bf k},\omega)=\Sigma^{(\rm h)}_{\rm 1L}({\bf k},\omega)
+\sigma_{x}\Sigma^{(\rm h)}_{\rm 1T}({\bf k},\omega)$ and $\Sigma^{(\rm h)}_{2}({\bf k},\omega)
=\Sigma^{(\rm h)}_{\rm 2L}({\bf k},\omega)+\sigma_{x}\Sigma^{(\rm h)}_{\rm 2T}({\bf k},\omega)$ in the particle-hole
and particle-particle channels, respectively, while the mean-field (MF) charge carrier normal Green's function
$g^{(0)}({\bf k},\omega)=g^{(0)}_{\rm L}({\bf k},\omega)+\sigma_{x}g^{(0)}_{\rm T}({\bf k},\omega)$, with the
corresponding longitudinal and transverse parts have been obtained as \cite{lan07,feng08},
\begin{subequations}
\begin{eqnarray}
g^{(0)}_{\rm L}({\bf k},\omega)&=&{1\over 2}\sum_{\alpha=1,2}{1\over \omega-\xi_{\alpha{\bf k}}},\\
g^{(0)}_{\rm T}({\bf k},\omega)&=&{1\over 2}\sum_{\alpha=1,2}(-1)^{\alpha+1}{1\over\omega-\xi_{\alpha{\bf k}}},
\end{eqnarray}
\end{subequations}
respectively, where $\alpha=1,2$, the MF charge carrier spectrum
$\xi_{\alpha{\bf k}}=Zt\chi_{1}\gamma_{\bf k}-Zt'\chi_{2}\gamma{'}_{\bf k}-\mu+(-1)^{\alpha+1}\chi_{\perp}
t_{\perp}({\bf k})$, the spin correlation function $\chi_{2}=\langle S_{ia}^{+}S_{i+\hat{\tau}a}^{-}\rangle$,
$\gamma_{\bf{k}}=(1/Z)\sum_{\hat{\eta}}{e^{i{\bf k}\cdot{\hat{\eta}}}}$,
$\gamma{'}_{\bf k}=(1/Z)\sum_{\hat{\tau}}{e^{i{\bf k}\cdot{\hat{\tau}}}}$, and $Z$ is the number of the nearest
neighbor or next nearest neighbor sites. However, in the bilayer coupling case, the more appropriate classification
is in terms of the normal and anomalous Green's functions within the basis of the bonding and antibonding components,
i.e., the full charge carrier normal and anomalous Green's functions can be rewritten in the bonding-antibonding
representation as,
\begin{subequations}
\begin{eqnarray}
g_{\nu}({\bf k},\omega)&=&g_{\rm L}({\bf k},\omega)+(-1)^{\nu+1}g_{\rm T}({\bf k},\omega),\\
\Im^{\dagger}_{\nu}({\bf k},\omega)&=&\Im^{\dagger}_{\rm L}({\bf k},\omega)
+(-1)^{\nu+1}\Im^{\dagger}_{\rm T}({\bf k},\omega),
\end{eqnarray}
\end{subequations}
respectively, where $\nu=1,2$, with $\nu=1$ ($\nu=2$) represents the corresponding bonding (antibonding) component,
then the bonding and antibonding components of the self-energies $\Sigma^{(\rm h)}_{1}({\bf k},\omega)$ and
$\Sigma^{(\rm h)}_{2}({\bf k},\omega)$ can be obtained from the spin bubble as \cite{lan07,feng08},
\begin{subequations}\label{self-energy}
\begin{eqnarray}
&&\Sigma^{(\rm h)}_{1\nu}({\bf k},i\omega_{n})=\Sigma^{(\rm h)}_{\rm 1L}({\bf k},i\omega_{n})
+(-1)^{\nu+1}\Sigma^{(\rm h)}_{\rm 1T}({\bf k},i\omega_{n})\nonumber\\
&&\hspace{0.5cm}={1\over 8N^{2}}\sum_{\bf pq}\sum_{\nu_{1}\nu_{2}\nu_{3}}\Lambda^{\nu\nu_{1}\nu_{2}\nu_{3}}_{{\bf p}+{\bf q}+{\bf k}}{1\over\beta}\sum_{ip_{m}}\nonumber\\
&&\hspace{0.8cm}\times g_{\nu_{1}}({\bf p}+{\bf k},ip_{m}+i\omega_{n})\Pi_{\nu_{2}\nu_{3}}({\bf p},{\bf q},ip_{m}),\\
&&\Sigma^{(\rm h)}_{2\nu}({\bf k},i\omega_{n})=\Sigma^{(\rm h)}_{\rm 2 L}({\bf k},i\omega_{n})
+(-1)^{\nu+1}\Sigma^{(\rm h)}_{\rm 2T}({\bf k},i\omega_{n})\nonumber\\
&&\hspace{0.5cm}={1\over 8N^{2}}\sum_{\bf pq}\sum_{\nu_{1}\nu_{2}\nu_{3}}\Lambda^{\nu\nu_{1}\nu_{2}\nu_{3}}_{{\bf p}+{\bf q}+{\bf k}}{1\over\beta}\sum_{ip_{m}}\nonumber\\
&&\hspace{0.8cm}\times \Im^{\dag}_{\nu_{1}} (-{\bf p}-{\bf k},-ip_{m}-i\omega_{n})\Pi_{\nu_{2}\nu_{3}}({\bf p},{\bf q},ip_{m}),~~~~~
\end{eqnarray}
\end{subequations}
respectively, with
$\Lambda^{\nu\nu_{1}\nu_{2}\nu_{3}}_{\bf k}=[1+(-1)^{\nu+\nu_{1}+\nu_{2}+\nu_{3}}][Z(t\gamma_{\bf k}-t'\gamma'_{\bf k})
+(-1)^{\nu+\nu_{3}}t_{\perp}({\bf k})]^{2}$, and the spin bubble,
\begin{eqnarray}
\Pi_{\nu_{2}\nu_{3}}({\bf p},{\bf q},ip_{m})&=&{1\over\beta}\sum_{iq_{m}}D^{(0)}_{\nu_{2}}({\bf q},iq_{m})\nonumber\\
&&\times D^{(0)}_{\nu_{3}}({\bf q}+{\bf p},iq_{m}+ip_{m}),
\end{eqnarray}
where the MF spin Green's functions
$D^{(0)}_{\nu}({\bf p},ip_{m})=B_{\nu{\bf p}}/[(ip_{m})^{2}-\omega^{2}_{\nu{\bf p}}]$, with the MF spin excitation
spectrum $\omega_{\nu{\bf p}}$ and function $B_{\nu{\bf p}}$ have been given in Ref. \cite{lan07,feng08}.

\begin{figure}[h!]
\includegraphics[scale=0.5]{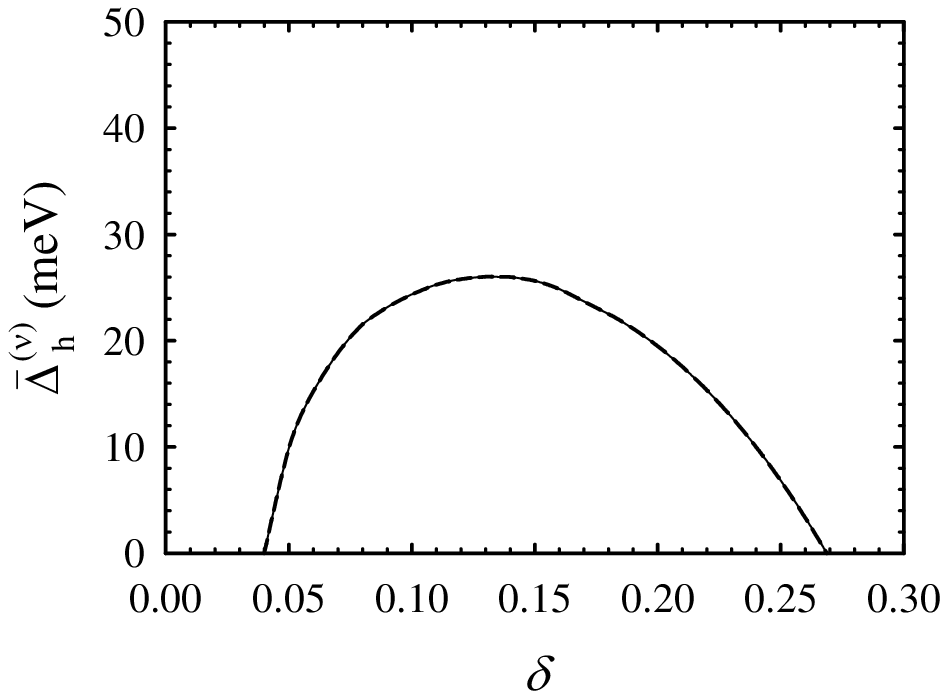}
\caption{The bonding (solid line) and antibonding (dashed line) components of the effective charge carrier pair gap
parameter as a function of doping for temperature $T=0.002J$ with parameters $t/J=2.5$, $t'/t=0.3$, $t_{\perp}/t=0.35$
and $J=110$ meV.  \label{fig1}}
\end{figure}

As in the single layer case \cite{feng0306}, the pairing force and charge carrier pair gap are incorporated into the
self-energy $\Sigma^{(\rm h)}_{2\nu}({\bf k},\omega)$, then it is called as the effective charge carrier pair gap
$\bar{\Delta}^{(\nu)}_{\rm h}({\bf k},\omega)=\Sigma^{(\rm h)}_{2\nu}({\bf k},\omega)$. On the other hand, the
self-energy $\Sigma^{(\rm h)}_{1\nu}({\bf k},\omega)$ renormalizes the MF charge carrier spectrum \cite{lan07,feng08}.
Moreover, $\Sigma^{(\rm h)}_{2\nu}({\bf k},\omega)$ is an even function of $\omega$, while
$\Sigma^{(\rm h)}_{1\nu}({\bf k},\omega)$ is not. For a convenience, $\Sigma^{(\rm h)}_{1\nu}({\bf k},\omega)$ can be
broken up into its symmetric and antisymmetric parts as, $\Sigma^{(\rm h)}_{1\nu}({\bf k},\omega)=
\Sigma^{(\rm h)}_{1\nu{\rm e}}({\bf k},\omega)+\omega\Sigma^{(\rm h)}_{1\nu{\rm o}}({\bf k},\omega)$, then both
$\Sigma^{(\rm h)}_{1\nu{\rm e}}({\bf k},\omega)$ and $\Sigma^{(\rm h)}_{1\nu{\rm o}}({\bf k},\omega)$ are an even
function of $\omega$. As in the conventional superconductors \cite{eliashberg60}, the retarded function
${\rm Re}\Sigma^{(\rm h)}_{1\nu{\rm e}}({\bf k},\omega)$ may be a constant, independent of (${\bf k},\omega$). It just
renormalizes the chemical potential, and therefore can be neglected. Now we define the charge carrier coherent weight
as $Z^{(\nu)-1}_{\rm hF}({\bf k},\omega)=1-{\rm Re}\Sigma^{(\rm h)}_{1\nu{\rm o}}({\bf k},\omega)$, and then in the
static limit approximation, i.e.,
$Z^{(\nu)-1}_{\rm hF}=1-{\rm Re}\Sigma^{(\rm h)}_{1\nu{\rm o}}({\bf k},\omega=0)\mid_{{\bf k}=[\pi,0]}$, and
$\bar{\Delta}^{(\nu)}_{\rm h}({\bf k})=\Sigma^{(\rm h)}_{2\nu}({\bf k},\omega=0)
=\Sigma^{(\rm h)}_{\rm 2 L}({\bf k},\omega=0)+(-1)^{\nu+1}\Sigma^{(\rm h)}_{\rm 2T}({\bf k},\omega=0)
=\bar{\Delta}_{\rm hL}({\bf k})+(-1)^{\nu+1}\bar{\Delta}_{\rm hT}({\bf k})$, with
$\bar{\Delta}_{\rm hL}({\bf k})=\bar{\Delta}_{\rm hL}\gamma^{(\rm d)}_{\bf k}$,
$\bar{\Delta}_{\rm hT}({\bf k})=\bar{\Delta}_{\rm hT}$, and
$\gamma^{(\rm d)}_{\bf k}=({\rm cos}k_{x}-{\rm cos}k_{y})/2$, we \cite{lan07,feng08} can obtain the full charge carrier
normal and anomalous Green's functions of the bilayer cuprate superconductors. In this case, with the help of these
full charge carrier normal and anomalous Green's functions, the self-energy $\Sigma^{(\rm h)}_{1\nu}({\bf k},\omega)$
and effective charge carrier pair gap $\bar{\Delta}^{(\nu)}_{\rm h}({\bf k})$ in Eq. (\ref{self-energy}) can be
evaluated explicitly as,
\begin{subequations}\label{self-energy-1}
\begin{eqnarray}
\Sigma^{(\rm h)}_{1\nu}({\bf k},\omega)&=&{1\over N^{2}}\sum_{{\bf pq}}\sum_{\nu_{1}\nu_{2}\nu_{3}}
\sum_{\sigma_{1}\sigma_{2}\sigma_{3}}\Lambda^{\nu\nu_{1}\nu_{2}\nu_{3}}_{{\bf p}+{\bf q}+{\bf k}}{B_{\nu_{2}{\bf q}}
B_{\nu_{3}{\bf q}+{\bf p}}\over 64\omega^{(\sigma_{2})}_{\nu_{2}{\bf q}}\omega^{(\sigma_{3})}_{\nu_{3}{\bf q+p}}}\nonumber\\
&\times& {A_{\sigma_{1}}^{(\nu_{1})}({\bf p}+{\bf k})F_{\sigma_{1}\sigma_{2}\sigma_{3}}^{\nu_{1}\nu_{2}\nu_{3}}({\bf p,q,k})\over \omega-E^{(\sigma_{1})}_{{\rm h}\nu_{1}{\bf p}+{\bf k}}-\omega^{(\sigma_{2})}_{\nu_{2}{\bf q}}
+\omega^{(\sigma_{3})}_{\nu_{3}{\bf q}+{\bf p}}},~~~~~~~\label{self-energy-1a}\\
\bar\Delta^{(\nu)}_{\rm h}({\bf k})&=&{1\over N^{2}}\sum_{\bf pq}\sum_{\nu_{1}\nu_{2}\nu_{3}}
\sum_{\sigma_{1}\sigma_{2}\sigma_{3}}\Lambda^{\nu\nu_{1}\nu_{2}\nu_{3}}_{{\bf p}+{\bf q}+{\bf k}}{B_{\nu_{2}{\bf q}}
B_{\nu_{3}{\bf q}+{\bf p}}\over 64\omega^{(\sigma_{2})}_{\nu_{2}{\bf q}}\omega^{(\sigma_{3})}_{\nu_{3}{\bf q}+{\bf p}}}\nonumber\\
&\times& {\bar\Delta_{\rm hZ}^{(\nu_{1})}({\bf p}+{\bf k})\over E^{(\sigma_{1})}_{{\rm h}\nu_{1}{\bf p}+{\bf k}}}
{F_{\sigma_{1}\sigma_{2}\sigma_{3}}^{\nu_{1}\nu_{2}\nu_{3}}({\bf p,q,k})\over
E^{(\sigma_{1})}_{{\rm h}\nu_{1}{\bf p}+{\bf k}}+\omega^{(\sigma_{2})}_{\nu_{2}{\bf q}}
-\omega^{(\sigma_{3})}_{\nu_{3}{\bf q}+{\bf p}}},~~~~~~~\label{self-energy-1b}
\end{eqnarray}
\end{subequations}
where $\sigma_{1},\sigma_{2},\sigma_{3}=1,2$, $F_{\sigma_{1}\sigma_{2}\sigma_{3}}^{\nu_{1}\nu_{2}\nu_{3}}({\bf p,q,k})=
Z^{(\nu_{1})}_{\rm hF}\{n_{\rm F}(E^{(\sigma_{1})}_{{\rm h}\nu_{1}{\bf p}+{\bf k}})
[n_{\rm B}(\omega^{(\sigma_{2})}_{\nu_{2}{\bf q}})-n_{\rm B}(\omega^{(\sigma_{3})}_{\nu_{3}{\bf q}+{\bf p}})]
+n_{\rm B}(\omega^{(\sigma_{3})}_{\nu_{3}{\bf q}+{\bf p}})[1+n_{\rm B}(\omega^{(\sigma_{2})}_{\nu_{2}{\bf q}})]\}$,
$A_{\sigma_{1}}^{(\nu_{1})}({\bf k})=1+\bar\xi_{\nu_{1}\bf k}/E^{(\sigma_{1})}_{{\rm h}\nu_{1}{\bf k}}$, with
$\omega^{(1)}_{\nu{\bf p}}=\omega_{{\nu}{\bf p}}$, $\omega^{(2)}_{\nu{\bf p}}=-\omega_{{\nu}{\bf p}}$,
$E^{(1)}_{{\rm h}\nu{\bf k}}=E_{{\rm h}\nu{\bf k}}$, $E^{(2)}_{{\rm h}\nu{\bf k}}=-E_{{\rm h}\nu{\bf k}}$, the
renormalized charge carrier excitation spectrum $\bar\xi_{\nu{\bf k}}=Z^{(\nu)}_{\rm hF}\xi_{\nu{\bf k}}$, the
renormalized charge carrier pair gap
$\bar\Delta_{\rm hZ}^{(\nu)}({\bf k})=Z^{(\nu)}_{\rm hF}\bar\Delta^{(\nu)}_{\rm h}({\bf k})$, the charge carrier
quasiparticle spectrum
$E_{{\rm h}\nu{\bf k}}=\sqrt{{\bar\xi}_{\nu{\bf k}}^{2}+\mid\bar{\Delta}_{\rm hZ}^{(\nu)}({\bf k})\mid^{2}}$, and
$n_{\rm B}(\omega)$ and $n_{\rm F}(E)$ are the boson and fermion distribution functions, respectively. In particular,
the equations $Z^{(\nu)-1}_{\rm hF}=1-{\rm Re}\Sigma^{(\rm h)}_{1\nu{\rm o}}({\bf k},\omega=0)\mid_{{\bf k}=[\pi,0]}$
and $\bar{\Delta}^{(\nu)}_{\rm h}({\bf k})=\Sigma^{(\rm h)}_{2\nu}({\bf k},\omega=0)$ have been solved
self-consistently in combination with other equations \cite{lan07,feng08}, then all order parameters and chemical
potential have been obtained by the self-consistent calculation. In Fig. \ref{fig1}, we plot the self-consistently
calculated result \cite{lan07} of the effective charge carrier pair gap parameter for the bonding
$\bar{\Delta}^{(1)}_{\rm h}$ (solid line) and antibonding $\bar{\Delta}^{(2)}_{\rm h}$ (dashed line) components
versus the doping concentration in temperature $T=0.002J$ with parameters $t/J=2.5$, $t'/t=0.3$, $t_{\perp}/t=0.35$,
and $J=110$ meV. It is shown clearly that the maximal $\bar{\Delta}^{(1)}_{\rm h}$ and $\bar{\Delta}^{(2)}_{\rm h}$
occur around the optimal doping, and then decrease in both the underdoped and the overdoped regimes. Moreover, both
the bonding and antibonding components of the effective charge carrier pair gap parameter have the same magnitude in
a given doping concentration \cite{lan07}, which implies the transverse part of the charge carrier pair gap
$\bar{\Delta}_{\rm hT}\approx 0$, and then
$\bar{\Delta}^{(1)}_{\rm h}\approx\bar{\Delta}^{(2)}_{\rm h}\approx\bar{\Delta}_{\rm hL}$. This result shows that
although there is a single electron {\it interlayer} coherent hopping (\ref{interlayer}) in the bilayer cuprate
superconductors, the coupling strength for the {\it interlayer} pairs vanishes, which reflects that within the
framework of the kinetic energy driven SC mechanism, the weak charge carrier interaction directly from the
{\it interlayer} coherent hopping (\ref{interlayer}) in the kinetic energy by exchanging spin excitations does not
provide any contribution to the charge carrier pair gap in the particle-particle channel, and then the transverse part
of the charge carrier pair gap $\bar{\Delta}_{\rm hT}\approx 0$. This is different from the strong charge carrier
interaction directly from the {\it intralayer} hopping in the kinetic energy by exchanging spin excitations, which
induces superconductivity in the particle-particle channel \cite{feng0306}, and then the charge carrier pair gap is
dominated by the corresponding longitudinal part, i.e.,
$\bar{\Delta}^{(1)}_{\rm h}\approx\bar{\Delta}^{(2)}_{\rm h}\approx\bar{\Delta}_{\rm hL}$. This result is also
consistent with the experimental results of the bilayer cuprate superconductor
Bi(Pb)$_{2}$Sr$_{2}$CaCu$_{2}$O$_{8+\delta}$ \cite{dfeng01}, where the SC gap separately for the bonding and
antibonding components has been measured, and it is found that both the antibonding and bonding components are
identical within the experimental uncertainties.

Now we discuss the interplay between the SC-gap and normal-state pseudogap in the bilayer cuprate superconductors. As
in the single layer case \cite{feng12}, the self-energy $\Sigma^{({\rm h})}_{1}({\bf k},\omega)$ in
Eq. (\ref{self-energy-1a}) in the particle-hole channel can also be rewritten approximately as,
\begin{eqnarray}\label{self-energy-2}
\Sigma^{(\rm h)}_{1}({\bf k},\omega)\approx {[2\bar{\Delta}_{\rm pg}({\bf k})]^{2}\over \omega+M_{\bf k}},
\end{eqnarray}
where $M_{\bf k}=M_{{\rm L}{\bf k}}+\sigma_{x}M_{{\rm T}{\bf k}}$ is the energy spectrum of
$\Sigma^{(\rm h)}_{1}({\bf k},\omega)$. As in the case of the effective charge carrier pair gap, the interaction force
and normal-state pseudogap have been incorporated into
$\bar{\Delta}_{\rm pg}({\bf k})=\bar{\Delta}_{\rm pgL}({\bf k})+\sigma_{x}\bar{\Delta}_{\rm pgT}({\bf k})$, and
therefore it is called as the effective normal-state pseudogap. In the bonding-antibonding representation, the
self-energy in Eq. (\ref{self-energy-2}) can be expressed as,
\begin{eqnarray}\label{self-energy-3}
\Sigma^{(\rm h)}_{1\nu}({\bf k},\omega)\approx {[2\bar{\Delta}^{(\nu)}_{\rm pg}({\bf k})]^{2}\over
\omega+M_{\nu{\bf k}}},
\end{eqnarray}
with $M_{\nu{\bf k}}=M_{{\rm L}{\bf k}}+(-1)^{\nu+1}M_{{\rm T}{\bf k}}$, and
$\bar{\Delta}^{(\nu)}_{\rm pg}({\bf k})=\bar{\Delta}_{\rm pgL}({\bf k})+(-1)^{\nu+1}\bar{\Delta}_{\rm pgT}({\bf k})$.
Substituting $\Sigma^{(\rm h)}_{1\nu}({\bf k},\omega)$ in Eq. (\ref{self-energy-3}) into Eq. (\ref{EliashbergEq}), the
full charge carrier normal and anomalous Green's functions can be obtained straightforwardly as,
\begin{widetext}
\begin{subequations}\label{fullGreenFunction}
\begin{eqnarray}
g_{\nu}({\bf k},\omega)&=&{1\over\omega-\xi_{\nu{\bf k}}-\Sigma^{(\rm h)}_{1\nu}({\bf k},\omega)
-[\bar{\Delta}^{(\nu)}_{\rm h}({\bf k})]^{2}/[\omega+\xi_{\nu{\bf k}}+\Sigma^{(\rm h)}_{1\nu}(-{\bf k},-\omega)]}
\nonumber\\
&=&{[U^{(\nu)}_{1{\rm h}{\bf k}}]^{2}\over\omega-E^{(\nu)}_{1{\rm h}{\bf k}}}+{[V^{(\nu)}_{1{\rm h}{\bf k}}]^{2}\over
\omega+E^{(\nu)}_{1{\rm h}{\bf k}}}+{[U^{(\nu)}_{2{\rm h}{\bf k}}]^{2}\over\omega-E^{(\nu)}_{2{\rm h}{\bf k}}}
+{[V^{(\nu)}_{2{\rm h}{\bf k}}]^{2}\over\omega+E^{(\nu)}_{2{\rm h}{\bf k}}},\\
\Im^{\dagger}_{\nu}({\bf k},\omega)&=&-{\bar{\Delta}^{(\nu)}_{\rm h}({\bf k})\over [\omega-\xi_{\nu{\bf k}}-
\Sigma^{(\rm h)}_{1\nu}({\bf k},\omega)][\omega+\xi_{\nu{\bf k}}+\Sigma^{(\rm h)}_{1\nu}(-{\bf k},-\omega)]-
[\bar{\Delta}^{(\nu)}_{\rm h}({\bf k})]^{2}}\nonumber\\
&=&-{\alpha^{(\nu)}_{1{\bf k}}\bar{\Delta}^{(\nu)}_{\rm h}({\bf k})\over 2E^{(\nu)}_{1{\rm h}{\bf k}}}\left({1\over
\omega-E^{(\nu)}_{1{\rm h}{\bf k}}}-{1\over\omega+E^{(\nu)}_{1{\rm h}{\bf k}}}\right)+{\alpha^{(\nu)}_{2{\bf k}}\bar{\Delta}^{(\nu)}_{\rm h}({\bf k})\over 2E^{(\nu)}_{2{\rm h}{\bf k}}}\left({1\over
\omega-E^{(\nu)}_{2{\rm h}{\bf k}}}-{1\over\omega+E^{(\nu)}_{2{\rm h}{\bf k}}}\right),
\end{eqnarray}
\end{subequations}
\end{widetext}
respectively, where $\alpha^{(\nu)}_{1{\bf k}}=\{[E^{(\nu)}_{1{\rm h}{\bf k}}]^{2}-M^{2}_{\nu{\bf k}}\}/
\{[E^{(\nu)}_{1{\rm h}{\bf k}}]^{2}-[E^{(\nu)}_{2{\rm h}{\bf k}}]^{2}\}$,
$\alpha^{(\nu)}_{2{\bf k}}=\{[E^{(\nu)}_{2{\rm h}{\bf k}}]^{2}-M^{2}_{\nu{\bf k}}\}/\{[E^{(\nu)}_{1{\rm h}{\bf k}}]^{2}
-[E^{(\nu)}_{2{\rm h}{\bf k}}]^{2}\}$,
$E^{(\nu)}_{1{\rm h}{\bf k}}=\sqrt{(\Omega_{\nu{\bf k}}+\Theta_{\nu{\bf k}})/2}$, and
$E^{(\nu)}_{2{\rm h}{\bf k}}=\sqrt{(\Omega_{\nu{\bf k}}-\Theta_{\nu{\bf k}})/2}$, with the kernel functions,
\begin{subequations}
\begin{eqnarray}
\Omega_{\nu{\bf k}}&=&\xi^{2}_{\nu{\bf k}}+M^{2}_{\nu{\bf k}}+8[\bar{\Delta}^{(\nu)}_{\rm pg}({\bf k})]^{2}
+[\bar{\Delta}^{(\nu)}_{\rm h}({\bf k})]^{2},\\
\Theta_{\nu{\bf k}}&=&\sqrt{(\xi^{2}_{\nu{\bf k}}-M^{2}_{\nu{\bf k}})\beta^{(\nu)}_{1{\bf k}}
+16[\bar{\Delta}^{(\nu)}_{\rm pg}({\bf k})]^{2}\beta^{(\nu)}_{2{\bf k}}+[\bar{\Delta}^{(\nu)}_{\rm h}({\bf k})]^{4}},\nonumber\\
\end{eqnarray}
\end{subequations}
where $\beta^{(\nu)}_{1{\bf k}}=\xi^{2}_{\nu{\bf k}}-M^{2}_{\nu{\bf k}}+2[\bar{\Delta}^{(\nu)}_{\rm h}({\bf k})]^{2}$,
$\beta^{(\nu)}_{2{\bf k}}=(\xi_{\nu{\bf k}}-M_{\nu{\bf k}})^{2}+[\bar{\Delta}^{(\nu)}_{\rm h}({\bf k})]^{2}$, while
the coherence factors,
\begin{subequations}\label{coherence-factors}
\begin{eqnarray}
(U^{(\nu)}_{1{\rm h}{\bf k}})^{2}&=&{1\over 2}\{\alpha^{(\nu)}_{1{\bf k}}[1+{\xi_{\nu{\bf k}}\over
E^{(\nu)}_{1{\rm h}{\bf k}}}]-\alpha^{(\nu)}_{3{\bf k}}[1+{M_{\nu{\bf k}}\over E^{(\nu)}_{1{\rm h}{\bf k}}}]\},\\
(V^{(\nu)}_{1{\rm h}{\bf k}})^{2}&=& {1\over 2}\{\alpha^{(\nu)}_{1{\bf k}}[1-{\xi_{\nu{\bf k}}\over
E^{(\nu)}_{1{\rm h}{\bf k}}}]-\alpha^{(\nu)}_{3{\bf k}}[1-{M_{\nu{\bf k}}\over E^{(\nu)}_{1{\rm h}{\bf k}}}]\},\\
(U^{(\nu)}_{2{\rm h}{\bf k}})^{2}&=&-{1\over 2}\{\alpha^{(\nu)}_{2{\bf k}}[1+{\xi_{\nu{\bf k}}\over
E^{(\nu)}_{2{\rm h}{\bf k}}}]-\alpha^{(\nu)}_{3{\bf k}}[1+{M_{\nu{\bf k}}\over E^{(\nu)}_{2{\rm h}{\bf k}}}]\},\\
(V^{(\nu)}_{2{\rm h}{\bf k}})^{2}&=& -{1\over 2}\{\alpha^{(\nu)}_{2{\bf k}}[1-{\xi_{\nu{\bf k}}\over
E^{(\nu)}_{2{\rm h}{\bf k}}}]-\alpha^{(\nu)}_{3{\bf k}}[1-{M_{\nu{\bf k}}\over E^{(\nu)}_{2{\rm h}{\bf k}}}]\},~~~~~~~
\end{eqnarray}
\end{subequations}
satisfy the sum rule: $[U^{(\nu)}_{1{\rm h}{\bf k}}]^{2}+[V^{(\nu)}_{1{\rm h}{\bf k}}]^{2}
+[U^{(\nu)}_{2{\rm h}{\bf k}}]^{2}+[V^{(\nu)}_{2{\rm h}{\bf k}}]^{2}=1$, with
$\alpha^{(\nu)}_{3{\bf k}}=[2\bar{\Delta}^{(\nu)}_{\rm pg}({\bf k})]^{2}/\{[E^{(\nu)}_{1{\rm h}{\bf k}}]^{2}-
[E^{(\nu)}_{2{\rm h}{\bf k}}]^{2}\}$, and then the corresponding effective normal-state pseudogap
$\bar{\Delta}^{(\nu)}_{\rm pg}({\bf k})$ and energy spectra $M_{\nu{\bf k}}$ can be obtained explicitly in terms of
the self-energies $\Sigma^{(\rm h)}_{1\nu}({\bf k},\omega)$ in Eq. (\ref{self-energy-1a}) as,
\begin{subequations}
\begin{eqnarray}
\bar{\Delta}^{(\nu)}_{\rm pg}({\bf k})&=& {L^{(\nu)}_{2}({\bf k})\over 2\sqrt{L^{(\nu)}_{1}({\bf k})}},
\label{pseudogap}\\
M_{\nu{\bf k}}&=& {L^{(\nu)}_{2}({\bf k})\over L^{(\nu)}_{1}({\bf k})},
\end{eqnarray}
\end{subequations}
where $L^{(\nu)}_{1}({\bf k})$ and $L^{(\nu)}_{2}({\bf k})$ are obtained from Eq. (\ref{self-energy-1a}) as,
\begin{subequations}
\begin{eqnarray}
L^{(\nu)}_{1}({\bf k})&=&{1\over N^{2}}\sum_{{\bf pq}}\sum_{\nu_{1}\nu_{2}\nu_{3}}
\sum_{\sigma_{1}\sigma_{2}\sigma_{3}}\Lambda^{\nu\nu_{1}\nu_{2}\nu_{3}}_{{\bf p}+{\bf q}+{\bf k}}{B_{\nu_{2}{\bf q}}
B_{\nu_{3}{\bf q}+{\bf p}}\over 64\omega^{(\sigma_{2})}_{\nu_{2}{\bf q}}\omega^{(\sigma_{3})}_{\nu_{3}{\bf q+p}}}\nonumber\\
&&\times {A_{\sigma_{1}}^{(\nu_{1})}({\bf p}+{\bf k})F_{\sigma_{1}\sigma_{2}\sigma_{3}}^{\nu_{1}\nu_{2}\nu_{3}}
({\bf p,q,k})\over (\omega^{(\sigma_{3})}_{\nu_{3}{\bf q}+{\bf p}}-E^{(\sigma_{1})}_{{\rm h}\nu_{1}{\bf p}
+{\bf k}}-\omega^{(\sigma_{2})}_{\nu_{2}{\bf q}})^{2}},~~~~~~~\\
L^{(\nu)}_{2}({\bf k})&=&{1\over N^{2}}\sum_{{\bf pq}}\sum_{\nu_{1}\nu_{2}\nu_{3}}
\sum_{\sigma_{1}\sigma_{2}\sigma_{3}}\Lambda^{\nu\nu_{1}\nu_{2}\nu_{3}}_{{\bf p}+{\bf q}+{\bf k}}{B_{\nu_{2}{\bf q}}
B_{\nu_{3}{\bf q}+{\bf p}}\over 64\omega^{(\sigma_{2})}_{\nu_{2}{\bf q}}\omega^{(\sigma_{3})}_{\nu_{3}{\bf q+p}}}\nonumber\\
&&\times {A_{\sigma_{1}}^{(\nu_{1})}({\bf p}+{\bf k})F_{\sigma_{1}\sigma_{2}\sigma_{3}}^{\nu_{1}\nu_{2}\nu_{3}}
({\bf p,q,k})\over \omega^{(\sigma_{3})}_{\nu_{3}{\bf q}+{\bf p}}-E^{(\sigma_{1})}_{{\rm h}\nu_{1}{\bf p}
+{\bf k}}-\omega^{(\sigma_{2})}_{\nu_{2}{\bf q}}}.
\end{eqnarray}
\end{subequations}
Now we obtain the effective normal-state pseudogap parameter from Eq. (\ref{pseudogap}) as,
\begin{eqnarray}
\bar{\Delta}^{(\nu)}_{\rm pg}&=&{1\over N}\sum_{\bf k}\bar{\Delta}^{(\nu)}_{\rm pg}({\bf k}),
\end{eqnarray}
where $\nu=1,2$, with $\bar{\Delta}^{(1)}_{\rm pg}$ and $\bar{\Delta}^{(2)}_{\rm pg}$ are the corresponding bonding and
antibonding components of the effective normal-state pseudogap parameter, respectively.

\begin{figure}[h!]
\includegraphics[scale=0.5]{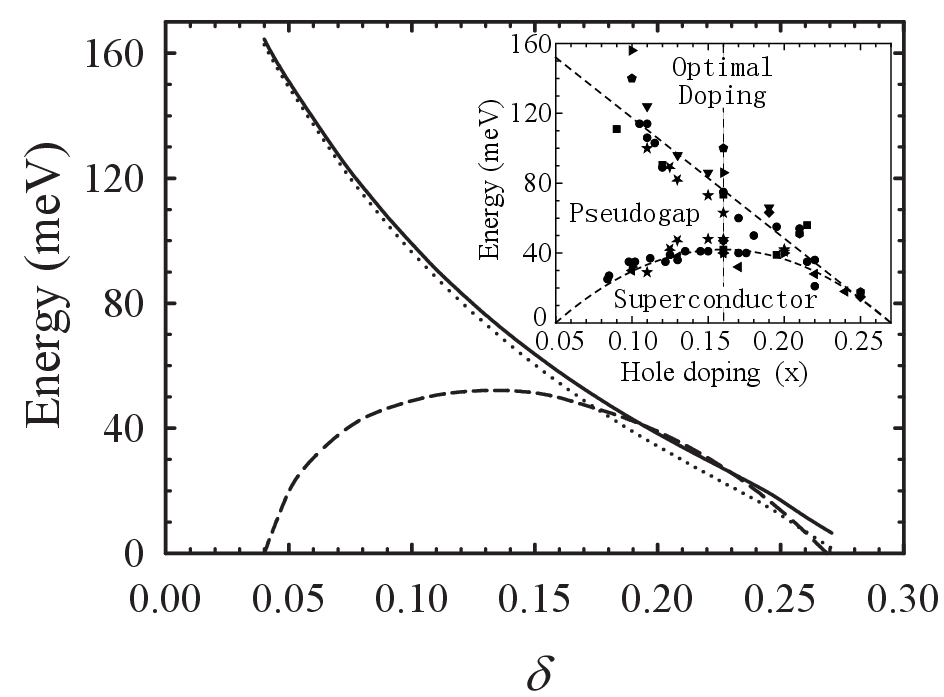}
\caption{The bonding (2$\bar{\Delta}^{(1)}_{\rm pg}$) (dotted line) and antibonding (2$\bar{\Delta}^{(2)}_{\rm pg}$)
(solid line) components of the effective normal-state pseudogap parameter, and the effective charge carrier pair gap
parameter ($2\bar{\Delta}_{\rm hL}$) (dashed line) as a function of doping for temperature $T=0.002J$ with
parameters $t/J=2.5$, $t'/t=0.3$, $t_{\perp}/t=0.35$ and $J=110$ meV. Inset: the experimental data observed on
different families of the cuprate superconductors taken from Ref. \onlinecite{Hufner08}. \label{fig2}}
\end{figure}

In Fig. \ref{fig2}, we plot the bonding (2$\bar{\Delta}^{(1)}_{\rm pg}$) (dotted line) and antibonding
(2$\bar{\Delta}^{(2)}_{\rm pg}$) (solid line) components of the effective normal-state pseudogap parameter, and the
effective charge carrier pair gap parameter ($2\bar{\Delta}_{\rm hL}$) (dashed line) as a function of doping for
$T=0.002J$ with $t/J=2.5$, $t'/t=0.3$, $t_{\perp}/t=0.35$ and $J=110$ meV in comparison with the corresponding
experimental data \cite{Hufner08} observed on different families of the cuprate superconductors (inset). Obviously,
both $\bar{\Delta}^{(1)}_{\rm pg}$ and $\bar{\Delta}^{(2)}_{\rm pg}$ have almost the same magnitude in a given doping
concentration, which implies the transverse part of the effective normal-state pseudogap parameter
$\bar{\Delta}_{\rm pgT}\approx 0$ and
$\bar{\Delta}^{(1)}_{\rm pg}\approx\bar{\Delta}^{(2)}_{\rm pg}\approx\bar{\Delta}_{\rm pgL}$, then the two-gap feature
observed on the bilayer cuprate superconductors \cite{Hufner08} is qualitatively reproduced. Moreover, the effective
normal-state pseudogap parameter $\bar{\Delta}_{\rm pgL}$ is much larger than the effective charge carrier pair gap
parameter $\bar{\Delta}_{\rm hL}$ in the underdoped regime, then it smoothly decreases with increasing the doping
concentration. In particular, both $\bar{\Delta}_{\rm pgL}$ and $\bar{\Delta}_{\rm hL}$ converge to the end of the SC
dome. The present result also shows that the weak charge carrier interaction directly from the {\it interlayer}
coherent hopping (\ref{interlayer}) in the kinetic energy by exchanging spin excitations does not provide the
contribution to the effective normal-state pseudogap in the particle-hole channel, and then the transverse part of the
effective normal-state pseudogap parameter $\bar{\Delta}_{\rm pgT}\approx 0$. On the other hand, the strong charge
carrier interaction directly from the {\it intralayer} hopping in the kinetic energy by exchanging spin excitations
therefore can induce the normal-state pseudogap in the particle-hole channel, and then the normal-state pseudogap is
dominated by the corresponding longitudinal part, i.e.,
$\bar{\Delta}^{(1)}_{\rm pg}\approx\bar{\Delta}^{(2)}_{\rm pg}\approx\bar{\Delta}_{\rm pgL}$, which is also consistent
with the experimental results of the bilayer cuprate superconductors \cite{Hufner08}, since only one normal-state
pseudogap is observed in the bilayer cuprate superconductors by using different measurement techniques \cite{Hufner08}.
In combination with the previous result of the single layer case \cite{feng12}, our present study suggests that the
single-layer model is sufficient for capturing the two-gap feature in cuprate superconductors.

It is well known that the many-body correlation and the related quasiparticle coherence in solids are closely related
to the electron self-energy. In particular, the positions of the low-energy quasiparticle peaks in the low-energy
excitation spectrum are determined by the electron self-energy. However, in the previous discussions of the electronic
structure based on the kinetic energy driven SC mechanism for both the single layer and bilayer cuprate superconductors
\cite{guo07,lan07,feng08}, the treatment of the charge carrier self-energy in the particle-hole channel is
oversimplified, i.e., in the static limit approximation, the charge carrier self-energy in the particle-hole channel
is replaced by the charge carrier coherent weight, then some subtle many-body effects from the normal-state pseudogap
is abandoned, which leads to that the peak-dip-hump structure in the low-energy excitation spectrum is absent from the
single layer cuprate superconductors \cite{guo07}, while the peak-dip-hump structure in the bilayer case is mainly
induced by BS \cite{lan07}. Recently, the electronic structure of the single layer cuprate superconductors has been
reexamined based on the kinetic energy driven SC mechanism by considering the normal-state pseudogap effect (then the
many-body correlation) beyond the previous static limit approximation for the charge carrier self-energy in the
particle-hole channel, and the result shows \cite{zhao12} that even in the single layer cuprate superconductors, there
is an obvious peak-dip-hump structure due to the presence of the normal-state pseudogap, in qualitative agreement with
the numerical result \cite{Ferrero09} based on the dynamical MF theory. In combination this result \cite{zhao12} for
the single layer cuprate superconductors and the previous result \cite{lan07} for the bilayer case, it suggests that
both the normal-state pseudogap and BS induce the peak-dip-hump structure in the bilayer cuprate superconductors,
however, the notable peak-dip-hump structure in the bilayer cuprate superconductors may be mainly dominated by BS.

The essential physics of the two-gap feature in the bilayer cuprate superconductors is the same as in the single layer
case \cite{feng12}, and can be attributed to the doping and temperature dependence of the charge carrier interactions
in the particle-hole and particle-particle channels directly from the kinetic energy by exchanging spin excitations.
Our present results also indicate that although BS due to the presence of the {\it interlayer} coherent hopping
(\ref{interlayer}) can play an important role in the form of the peak-dip-hump structure around the antinodal point
\cite{lan07,feng08}, it may have not an impact on the overall global feature for the SC gap and normal-state pseudogap
parameters. This follows a fact that BS is maximum around the antinodal point, and it vanishes along the nodal
direction. As an result, this momentum dependence of BS has an impact on the momentum dependence of the peak-dip-hump
structure, while it does no has an effect on the momentum independence of the SC gap and normal-state pseudogap
parameters. Furthermore, in the present bilayer case, we have also calculated the doping dependence of the coupling
strength $V_{\rm eff}$, and the result shows that as in the single layer case \cite{feng12}, the coupling strength
$V_{\rm eff}$ smoothly decreases upon increasing the doping concentration from a strong-coupling case in the underdoped
regime to a weak-coupling side in the overdoped regime. Since the charge carrier interactions in both the particle-hole
and particle-particle channels are mediated by the same spin excitations as shown in Eq. (\ref{self-energy}), therefore
all these charge carrier interactions are controlled by the same magnetic interaction $J$. In this sense, both the
normal-state pseudogap and SC gap in the phase diagram of the bilayer cuprate superconductors are dominated by one
energy scale. This is why both $\bar{\Delta}_{\rm pgT}\approx 0$ (then
$\bar{\Delta}^{(1)}_{\rm pg}\approx\bar{\Delta}^{(2)}_{\rm pg}\approx\bar{\Delta}_{\rm pgL}$) and
$\bar{\Delta}_{\rm hT}\approx 0$ (then
$\bar{\Delta}^{(1)}_{\rm h}\approx\bar{\Delta}^{(2)}_{\rm h}\approx\bar{\Delta}_{\rm hL}$) simultaneously in the
bilayer cuprate superconductors, and then the two-gap behavior is a universal feature for the single layer and bilayer
cuprate superconductors.

In conclusion, we have discussed the interplay between the SC gap and normal-state pseudogap in the bilayer cuprate
superconductors based on the framework of the kinetic energy driven SC mechanism. Our results show that the
single-layer model is sufficient for capturing the two-gap feature in cuprate superconductors. The weak charge carrier
interaction directly from the {\it interlayer} coherent hopping (\ref{interlayer}) in the kinetic energy by exchanging
spin excitations does not provide the contribution to the normal-state pseudogap in the particle-hole channel and
charge carrier pair gap in the particle-particle channel, which leads to that the transverse parts of the effective
normal-state pseudogap parameter $\bar{\Delta}_{\rm pgT}\approx 0$ and effective charge carrier pair gap parameter
$\bar{\Delta}_{\rm hT}\approx 0$ simultaneously, while only the strong charge carrier interaction directly from the
{\it intralayer} hopping in the kinetic energy by exchanging spin excitations therefore induces the normal-state
pseudogap in the particle-hole channel and charge carrier pair gap in the particle-particle channel, and then the
normal-state pseudogap and charge carrier pair gap are dominated by the corresponding longitudinal parts, i.e.,
$\bar{\Delta}^{(1)}_{\rm pg}\approx\bar{\Delta}^{(2)}_{\rm pg}\approx\bar{\Delta}_{\rm pgL}$
and $\bar{\Delta}^{(1)}_{\rm h}\approx\bar{\Delta}^{(2)}_{\rm h}\approx\bar{\Delta}_{\rm hL}$.

\acknowledgments

The authors would like to thank Dr. Huaisong Zhao for helpful discussions. The part of the numerical calculations is
performed by using the Siyuan clusters. YL is supported by the National Natural Science Foundation of China (NSFC)
under Grant No. 11004084, JQ is supported by NSFC under Grant No. 11004006, and SF is supported by NSFC under Grant
Nos. 11074023 and 11274044, and the funds from the Ministry of Science and Technology of China under Grant Nos.
2011CB921700 and 2012CB821403.


\begin{thebibliography}{00}

\bibitem{schrieffer83} J. R. Schrieffer, \emph{Theory of Superconductivity} (Addison-Wesley, San Francisco, 1964).

\bibitem{Hufner08} See, e.g., S. H\"ufner, M. A. Hossain, A. Damascelli, and G. A. Sawatzky, Rep. Prog. Phys. {\bf 71},
062501 (2008), and references therein.

\bibitem{Timusk99} See, e.g., Tom Timusk and Bryan Statt, Rep. Prog. Phys. {\bf 62}, 61 (1999), and references therein.

\bibitem{damascelli03} See, e.g., A. Damascelli, Z. Hussain, and Z.-X. Shen, Rev. Mod. Phys. {\bf 75}, 473 (2003), and
references therein.

\bibitem{campuzano04} See, e.g., J. C. Campuzano, M. R. Norman, and M. Randeira, in {\it Physics of Superconductors},
vol. II, edited by K. Bennemann and J. Ketterson (Springer, Berlin Heidelberg New York, 2004), p. 167, and references
therein.

\bibitem{campuzano03} J. C. Campuzano, H. Ding, M. R. Norman, M. Randeira, A. F. Bellman, T. Yokoya, T. Takahashi, H.
Katayama-Yoshida, T. Mochiku, and K. Kadowaki, Phys. Rev. B {\bf 53}, R14737 (1996); H. Matsui, T. Sato, T. Takahashi,
S.-C. Wang, H.-B. Yang, H. Ding, T. Fujii, T. Watanabe, and A. Matsuda, Phys. Rev. Lett. {\bf 90}, 217002 (2003).

\bibitem{dfeng01} D. L. Feng, N. P. Armitage, D. H. Lu, A. Damascelli, J. P. Hu, P. Bogdanov, A. Lanzara, F. Ronning,
K. M. Shen, H. Eisaki, C. Kim, Z.-X. Shen, J.-i. Shimoyama, and K. Kishio, Phys. Rev. Lett. {\bf 86}, 5550 (2001);
A. A. Kordyuk, S. V. Borisenko, M. Knupfer, and J. Fink, Phys. Rev. B {\bf 67}, 064504 (2003); Y.-D. Chuang, A. D.
Gromko, A. Fedorov, Y. Aiura, K. Oka, Y. Ando, H. Eisaki, S. I. Uchida, and D. S. Dessau, Phys. Rev. Lett. {\bf 87}, 
117002 (2001).

\bibitem{lan07} Yu Lan, Jihong Qin, and Shiping Feng, Phys. Rev. B {\bf 75}, 134513 (2007); Yu Lan, Jihong Qin, and
Shiping Feng, Phys. Rev. B {\bf 76}, 014533 (2007).

\bibitem{kordyuk02} A. A. Kordyuk, S. V. Borisenko, T. K. Kim, K. A. Nenkov, M. Knupfer, J. Fink, M. S. Golden, H.
Berger, and R. Follath, Phys. Rev. Lett. {\bf 89}, 077003 (2002).

\bibitem{borisenko03} S. V. Borisenko, A. A. Kordyuk, T. K. Kim, A. Koitzsch, M. Knupfer, J. Fink, M. S. Golden, M.
Eschrig, H. Berger, and R. Follath, Phys. Rev. Lett. {\bf 90}, 207001 (2003); S. V. Borisenko, A. A. Kordyuk, T. K.
Kim, S. Legner, K. A. Nenkov, M. Knupfer, M.S. Golden, J. Fink, H. Berger, and R. Follath, Phys. Rev. B {\bf 66}, 
140509 (2002).

\bibitem{feng0306} Shiping Feng, Phys. Rev. B {\bf 68}, 184501 (2003); Shiping Feng, Tianxing Ma, and Huaiming Guo,
Physica C {\bf 436}, 14 (2006).

\bibitem{feng12} Shiping Feng, Huaisong Zhao, and Zheyu Huang, Phys. Rev. B {\bf 85}, 054509 (2012).

\bibitem{anderson87} P. W. Anderson, Science {\bf 235}, 1196 (1987).

\bibitem{chakarvarty95} O. K. Anderson, A. I. Liechtenstein, O. Jepsen, and F. Paulsen, J. Phys. Chem. Solids {\bf 56}, 
1573 (1995); A. I. Liechtenstein, O. Gunnarsson, O. K. Anderson, and R. M. Martin, Phys. Rev. B {\bf 54}, 12505 (1996).

\bibitem{feng04} Shiping Feng, Jihong Qin, and Tianxing Ma, J. Phys.: Condens. Matter {\bf 16}, 343 (2004).

\bibitem{feng08} See, e.g., the review, Shiping Feng, Huaiming Guo, Yu Lan, and Li Cheng, Int. J. Mod. Phys. B
{\bf 22}, 3757 (2008).

\bibitem{eliashberg60} G. M. Eliashberg, Sov. Phys. JETP {\bf 11}, 696 (1960); D. J. Scalapino, J. R. Schrieffer, and
J. W. Wilkins, Phys. Rev. {\bf 148}, 263 (1966).

\bibitem{guo07} Huaiming Guo and Shiping Feng, Phys. Lett. A {\bf 361}, 382 (2007).

\bibitem{zhao12} Huaisong Zhao, L\"ulin Kuang, and Shiping Feng, Physica C {\bf 483}, 225 (2012); Huaisong Zhao and
Shiping Feng, unpublished.

\bibitem{Ferrero09} Michel Ferrero, Pablo S. Cornaglia, Lorenzo De Leo, Olivier Parcollet, Gabriel Kotliar, and Antoine
Georges, Phys. Rev. B {\bf 80}, 064501 (2009).

\end{thebibliography}
\end{document}